\documentclass[11pt,epsfig]{article}
\usepackage{graphicx}

 1
\font\elevenrm=cmr10 scaled\magstep 1

\renewenvironment{thebibliography}[1]
 { \elevenrm
   \begin{list}{\arabic{enumi}.}
    {\usecounter{enumi}     \setlength{\parsep}{0pt}
     \setlength{\itemsep}{3pt} \settowidth{\labelwidth}{#1.}
     \sloppy
    }}{\end{list}}

\begin{document}
\title{{\bf On Exotic Systems of Baryons in Chiral Soliton Models}\\}
\author{Vladimir Kopeliovich$^{a,b}$\footnote{{\bf e-mail}: kopelio@inr.ru}
\\
\small{\em a) Institute for Nuclear Research of RAS, Moscow 117312, Russia}
\\
\small{\em b) Nordita} 
\\
}

\date{}
\maketitle

DOI: 10.1016/0370-2693(91)90821-7
\begin{abstract}
The role of  zero mode quantum corrections to the energy of baryonic systems with exotic quantum numbers (strangeness) 
is discussed. A simple expression for the contribution depending on strange inertia is obtained in the $SU(3)-$collective 
coordinate quantization approach, and it is shown that this correction stabilizes the systems the stronger the greater 
their baryon number is. Furthemore, systems are considered which could be interpreted in the quark model language as
containing additional $q\bar q-$pairs. It is argued that a strange skyrmion crystal should have additional binding in 
comparison with the $SU(2)-$quantized neutral crystal.
\end{abstract}

\section{Introduction}
One of the most characteristic predictions of chiral soliton models, such as the Skyrme model, is the possible existence
of the $SU(3)-$multiplets of dibaryons, tribaryons, etc., whose strange components could be stable relative to strong
interactions \cite{1} - \cite{5}. These predictions followed the paper by Jaffe \cite{6} where the existence of a strange 
dibaryon, the $H-$particle, was predicted in the MIT bag model. The experimantal checking of these predictions would be
very important since it would provide evidence for the whole concept of baryons as solitons of some effcetive lagrangian, 
or lead to  substantial modifications of these models.

The purpose of present note is to clear up some general properties of exotic baryon systems in the framework of chiral
soliton models and the collective coordinates quantization picture which have not been well understood in previous
investigations. We shall use here the rigid rotator model criticized, e.g. in \cite{7}, so the results obtained should
be considered as a starting point for more elaborate investigations. Arguments based on a very simple expression for the
quantum correction due to rotation in a "strange" direction show that this correction has the property of stabilizing 
the baryon systems more strongly than than the quantum corrections due to other zero modes, the value of the binding
energy being increased with increasing baryon number $B$ of the system.

Below we shall consider objects which can be interpreted as being built of a minimal number of quarks (valence quarks 
only) as well as those which contain additional quark-antiquark pairs. The latter may have positive strangeness as well 
as a ratio of strangeness to baryon number $|S/B| > 3$, for some of possible multiplets. The 'price' to be paid for addition
of each $q\bar q$-pair is estimated, and it is shown that this price is modest. Finally, arguments are given that in 
the case of strange skyrmion crystals quantum correction due to zero modes stabilize them more strongly than $SU(2)-$quantized
skyrmion crystals (neutron crystals).

\section{$SU(3)$ - quantum corrections for nonexotic and exotic systems}
Our consideration is based on the well known expression for the energy of baryon systems obtained from the $SU(2)-$solitons
by means of rotations in the space $SU(3)-$collective coordinates \cite{1} - \cite{5}. This energy consists of the
classical mass of the soliton $M_{cl}$ and quantum corrections due to zero modes $E_{rot}$ as well as nonzero modes
(vibration, breathing, etc) which are difficult to calculate and by this reason are usually omitted:
$$E= M_{cl} + E_{rot} = $$
$$= M_{cl} + {1\over \lambda_s}\left[C_2(SU(3)) - {3Y'^2\over 4} - N(N+1)\right] + {N(N+1)\over \lambda_r} +... \eqno(1) $$

$M_{cl}$ is bound, at least for not very large baryon numbers $B$ (the masses of toroidal bound skyrmions with $B=2,\,3,\,4$
have been calculated at first in ref. \cite{8}). $C_2(SU(3))=[p^2+q^2+pq + 3(p+q)]/3$ is the $SU(3)$ Casimir operator, $p$ 
and $q$ being the numbers of upper (lower) indeces in the spinor describing the $SU(3)$ multiplet under consideration, 
$Y'$ and $N$ are the right hypercharge and isospin (the mistake in the interpretation of $N$ in \cite{3} has been corrected in
\cite{5}), $\lambda_s$ and $\lambda_r$ are the moments of inertia characterizing the system, they differ by a factor of two
from their standard definition (accepted in classical mechanics, e.g.). The terms depending on the orbital momentum, like 
$J(J+1)/\Lambda_r$ will be discussed later. The expresiion $(1)$ was obtained for systems posessing the generalized axial
symmetry \cite{8,5}.  As it was recently established in \cite{9} for systems with baryon numbers $B = 3 - 6$ at least, 
there exist configurations of more complicated form than the torus-like one, which have even lower energy. However, we
expect that some changes in the spatial form of the solitons will not influence the overall structure of the expression $(1)$,
so, our our arguments hold also for this case.

Let us consider the $SU(3)-$multiplets for which the right hypercharge $Y'$ is the largest in the multiplet (we call them the
"minimal" multiplets):
$$ Y' = {p+2q \over 3}, \eqno(2) $$ 

From the quantization condition \cite{10} it follows that $Y' = N_cB/3 $, but here we shall restrict ourselves to the case
of the number of colors $N_c=3$. Since the right isospin $N = p/3$, it is easy to establish that the coefficient of $1/\lambda_s$
in $(1)$ is equal to $(p+2q)/2 = 3B/2$.  So, for the whole family of minimal multiplets with fixed baryon number and $N$ varying 
from $0$ or $1/2$ up to $3B/2$ we obtain the universal relation
$$E_{rot} = {3B\over 2\lambda_s} +{N(N+1)\over \lambda_r} +{J(J+1)\over \Lambda_r} +$$
$$+ J_z^2 \left[{1\over B^2\lambda_z} - {1\over B^2\lambda_r} -{1\over \Lambda_r} \right]. \eqno (3) $$

It was already noted in \cite{1} that the first coefficient in $(3)$ is the same for the octet and decuplet of baryons (hyperons). 
Therefore, measurements of their masses cannot help in cross-checking the $\lambda_s$. The same holds also for families of
baryonic systems with arbitrary baryon numbers. These families consist of $\overline{10}-,\; 27-,\;35-,\;28-$plets for $B=2$,
of $\overline{35}-,\; 64-,\;81-,\;80-$  and $55-$plets for $B=3$, etc \cite{11}.

For toroidal few-baryon systems considered in \cite{4,8} $\lambda_s(B)$ increases proportional to the baryon number of the 
system, or even faster. The same holds also for $\lambda_r(B)$. The orbital inertia $\Lambda_r(B)$ is considerably larger than
$\lambda_s$ and $\lambda_r$, it grows with increasing $B$ faster than $B^2$. By this reason the depending on  $\Lambda_r$
contribution in $(3)$ is smaller than the first two terms, if $J$ is not very large, and does not play a crucial role
\footnote{Even for a rotating neutron star the energy of the orbital rotation is considerably smaller than the sum of the 
rotational energies of individual skyrmions (neutrons) if the angular velocity of rotation is not very large (period 
$T>10^{-3}$}. in the binding of the system. It follows immediately from $(3)$ that the energy (due to zero modes) per unit 
baryon number decreases with increasing $B$ and goes to zero for rotations in the "strange" direction. The energy due to 
isotopic rotations decreases like $1/B^2$ for the multiplets with the smallest value of $N$ ($1/2$ or $1$), but is 
approximately constant for $N=N_{max} = 3B/2$. 

The contribution to the binding energy per unit baryon number for the systems with smallest $N$ increases with increasing $B$ 
and approaches the value
$${\epsilon\over B} \simeq {3\over 2\lambda_s(B=1)} + {3\over 4 \lambda_r(B=1)}. \eqno (4) $$
Since usually $\lambda_s$ is a few times smaller than $\lambda_r$, "strange" rotational energy binds baryonic systems more
strongly than isotopic rotations \cite{1,4,5}. The numerical results for the binding energy depend on the values of the
model parameters. It follows from $(4)$ that the asymptotic relative binding energy per baryon equals to 0.33 in the Skyrme
model \cite{1,4} and to 0.46 in the model with explicit scalar meson \cite{5}. These values seem to be too large, but they
are smaller in the case of few-baryon systems. The relative binding energy of quantized dibaryons (i.e. divided by the total
energy of possible final states) is less sensitive to the type of the model and to the values of the parameters, and in many 
cases of interest is close to $0.2$.

Let us go to the "nonminimal" representations for which the largest possible hypercharge of the system is larger than their
baryon numbers. Such multiplets have to contain components with positive strangeness. In the quark model language these systems
could contain some number $m$ of quark-antiquark pairs. We have now
$$ {p+2q\over 3} = B+m $$ and
$$ Y' ={p+2q \over 3} - m, $$ 
if we assume that in a nonminimal representation the number of indices is increased by $m$ for both $p$ and $q$ in comparison
with corresponding minimal representation. Of course, this is not the only possibility. In this case the right isospin 
$N=(p+m)/2 = N_0 +m$, and an elementary calculation yields
$$ C_2(SU(3)) - N(N+1) - {3B^2\over 4}\; = \;{3B\over 2} +m\left({3B\over 2} +m -N +1\right). \eqno(5)$$
So, the increase in energy of the system due to the addition of $m\;\, q\bar q $ pairs contains terms linear in $m$ as well
as quadratic ones. At fixed $B$ and $m$ expression $(5)$ decreases with increasing $N$ and is minimal for 
$$ N_{max} = {3B\over 2} +m, $$
the second term in the above formula being equal to $m$. Taking into account the contribution to the energy due to isotopic
rotations we obtain for the total increase of the energy due to the addition of $m$ quark-antiquark pairs:
$$\delta E_{rot} = m\left[{3B/2 +1 +m -N\over \lambda_s} + {2N+1 -m\over \lambda_r}\right]. \eqno(6) $$
For an octet of baryons, e.g., $\delta E_{rot} =2/\lambda_s + 3/\lambda_r$, at $m=1$ \footnote{In this case for starting
configuration $(p_0,q_0) =(1,1)$, $N_0=1/2$, and after addition of one $q\bar q-$pair we have $(p,q)=(2,2)$, $N=3/2$.}.
The contribution of the rotation into the "strange" direction decreases with increasing $N$ down to $m/\lambda_s$, 
so, when strange quarks are "dissolved" in multiplets of higher dimension (in $p$ or $N$) the energy necessary
for this "dissolving" decreases. However, the energy of the isotopic rotations increases.

There are also other ways to go to nonminimal representations by means of asymmetric increase of $p$ and $q$. E.g., 
one can increase separately $p$ or $q$  by $3m$ which will correspond to the addition of $m$ or $2m$ quark-antiquark pairts.
In the first case $N=(m+p)/2$, in the second one $N=p/2 + m$, the expression for $\delta E$ remains the same with 
the substitution $m\to 2m$ in the latter case. In the Skyrme model with the parameters $F_\pi = 108\,MeV$, $e=4.84$ 
we have  $\lambda_s^{-1} \simeq 0.3\,GeV$ and  $\lambda_r^{-1} \simeq 0.1\,GeV$, and the energy surplus for $m=1$ equals 
$$\delta E_{rot} = \left[0.3\left({3B\over 2} +2 -N\right) + 0.2 N\right].  $$
In the model with an additional scalar meson \cite{5} it is larger by a factor $\sim 1.5$. So, $\delta E$ has the 
desired order of magnitude because it is expected that such states should be above the threshold fot the 
decay due to strong interactions.

\section{Quantum corrections in case of skyrmion crystals } 
The binding energy of the extended skyrmion crystal consists of two parts: the classical binding energy and the
binding energy due to quantum corrections \cite{12} - \cite{17}. The first one depends on the particular symmetry 
of the crystal. We shall concentrate now on the contribution of the quantum corrections to the binding energy
and show that there is a principal distinction between the contribution due to the rotations in the "strange" and 
"nonstrange" direction.

Let us recall the arguments of ref. \cite{12} which have led to the conclusion that there is an important contribution
of the quantum corrections to the binding energy of the crystal. The basic assumption is that flavor rotations of 
the whole crystal can be described by one and the same set of collective coordinates, i.e. the crystal is coherently 
rotated in flavor space. It follows immediately that the corresponding moment of inertia of the crystal $\Lambda = n\lambda$,
where $n$ is the number of unit cells in the crystal, and $\lambda$ is the moment of inertia of the unit cell. The 
rotational energy of the whole crystal $E_{rot}= T(T+1)/n\lambda$ should be compared with $nt(t+1)/\lambda$ where $t=1/2$
is the isospin of the unit cell. For the neutron crystal $T=nt=n/2$, and for large $n$ we have
$$E_{rot}= nt^2/\lambda\,<\,  nt(t+1)/\lambda, \eqno(7) $$
i.e. the effect of binding arises. Its physical sources are the quantum fluctuations of the isospin momentum of the
free neutron which make $\vec t^2$ three times greater than $t_z^2$. These fluctuations are suppressed inside the
crystal, and for the whole crystal they are negligible. It was assumed in the above arguments that the inertia of the
unit cell does not differ too much from that of the free neutron.

If we consider now the $SU(3)$ flavor rotations we obtain under the same assumptions for the energy of the crystal:
$$E_{rot} = {1\over n\lambda_s}\left[C_2(SU(3))[P,Q] -{3B^2\over 4}-N^{tot}(N^{tot} +1)\right]+
{N^{tot}(N^{tot}+1)\over n\lambda_r},
\eqno(8) $$
$B=n$, which should be compared with the  analogous expression for $n$ unit cells. $P$ and $Q$ are the numbers of indeces
in the spinor describing the crystal in the $SU(3)$ space, $N^{tot}$ is the corresponding right isospin, $N^{tot} = P/2$.
The simplest and probably most natural possibility is to assume that $P=np$ (by analogy with the neutron crystal).
In this case we obtain 
$$E_{rot} = {3B\over 2\,n\lambda_s} +{n^2\over 4\,n\lambda_r}\, =\, {3\over 2\lambda_s} + {n\over 4 \lambda_r}. \eqno(9)$$
This expression is quite similar to the expression for $E_{rot}$ in the case of few-baryon systems \footnote{The first
term in Eq. $(9)$ should be compared with corresponding contribution for $B=n$ separate cells (baryons), equal to
$3n/2\lambda_s$}. The distinctions are
that now $n\lambda$ enters instead of $\lambda(B)$, $n/2$ instead of $N$, and the orbital rotation energy is absent (for 
the crystal at rest). So, due to the cancellation of the quadratic terms in $P,\,Q$ we have obtained that the contribution
of the first term in $(9)$ to the energy of the crystal is constant and equal to $3/2\lambda_s$. This is the main difference
between "strange" and isotopical rotations. As in the case of few-baryon systems, this term is the same for all $SU(3)$
representations $(P,Q)$ which satisfy the relation $P+2Q = 3n$.

The discussion of the astrophysical aspects of the possible existence of strange matter is beyond the scope of the present
paper. Observational results imply that there is a place for dark baryonic matter in the Universe, although arguments
were given that the strange matter should have evaporated at the early stages of the evolution of the Universe \cite{18}.
However, the possibility of the existence of strange matter does not seem to be excluded completely.
\\
\section{Discussion of some general problems} 
We have shown that in the collective coordinate quantization approach a transparent expression for the quantum correction to
the energy of the baryon system can be obtained, illustrating that rotations into "strange" direction stabilize few-baryon systems
as well as skyrmion crystals even more than isotopic rotations. The stabilizing property of zero modes quantum corrections
seems to be a simple consequence of a natural property of solitons: their geometrical sizes and moments of inertia increase 
with increasing baryon number.

The fact that the quantum corrections due to "strange" rotations decrease with increasing baryon number of the system may 
also mean that the whole approach of the $SU(3)-$ collective coordinates quantization becomes more selfconsistent with
increasing $B$. Indeed, several attempts to to describe the $B=1$ hyperon spectrum \cite{10,19,20} have met problems since
the strange inertia $\lambda_s$ is especially small for $B=1$, and therefore the quantum correction is especially large in 
this case. As shown above, the relative value of this correction drops as $1/B$, so the problems in the $B=1$ sector disappear 
for systems with large $B$. However, if this argument is correct, the same result should be obtained in other versions of 
quantization, e.g. in that of Callan and Klebanov. In other words, it may turn out that systems with large $B$ are in some 
sense simpler objects than system with $B=1$. There remain, however, some effects which role has not been clarified till now.

First of all, the vibrational corrections remain to be calculated. They should increase the energy of the system, as well as
of the whole crystal. For the system with $B=1$ they have been analyzed in \cite{21,22}. An elegant method for estimating 
them has been proposed in \cite{23}, and for the case of the skyrmion crystal in \cite{15}.

The other problem is the applicability of the soliton approach with increasing baryon number. It is clear that under usual 
conditions the soliton (skyrmion) cannot be too large, e.g. with baryon number of the order of the Avogadro number. But it
is unclear what parameter defines the maximal baryon number of the soliton.

We have not included in our consideration the meson mass terms in the effective lagrangian of the model. According to our 
previous experience \cite{3} - \cite{5} the mass terms do not play a crucial role in the binding of quantized skyrmions, although 
they define the mass splittings inside the multiplets, which happened to be too small in comparison with experimental data, 
especially in Syracuse model \cite{5}. The mass terms induce also the mass splittings of the quantized skyrmion crystals.
\\

The author is greatly indebted to Andreas Wirzba for many useful discussions and reading of the manuscript, and to Andy Jackson
for enlightning discussions of some points of principle.\\

\end{document}